\documentclass[12pt]{amsart}
\usepackage{amssymb}
\usepackage{amsmath}
\usepackage{graphicx}
\usepackage{subfigure}
\usepackage{hyperref}
\usepackage{cite}
\usepackage{color}
\usepackage{framed}
\definecolor{shadecolor}{rgb}{.90,.95,1}
\numberwithin{equation}{section}

\newcommand{\norm}[1]{\lVert#1\rVert}
\newcommand{\paren}[1]{\left(#1\right)}
\newcommand{\bracket}[1]{\left[#1\right]}

\newcommand{\p}{\partial}

\newcommand{\bigo}{\mathrm{O}}

\newcommand{\dx}{\partial_{x} }

\newcommand{\sinc}{\mathrm{sinc}}

\newcommand{\cstar}{{c_\star}}

\newcommand{\inner}[2]{\left\langle #1,#2\right\rangle}

\newcommand{\dc}{\partial_{c}}

\newcommand{\dy}{\partial_{y}}

\newcommand{\R}{\mathbb{R}}

\newcommand{\calN}{\mathcal{N}}
\newcommand{\calI}{\mathcal{I}}

\newcommand{\calE}{\mathcal{E}}

\thanks{The first author was partially supported by an IBM Junior
  Faculty Development Award through the University of North Carolina.
  The second author was partially supported by a grant from the Simons Foundation.  The third author was supported by NSERC, and his contribution to
  this work was completed under the NSF PIRE grant OISE-0967140 and
  the DOE grant DE-SC0002085. }

\begin{document}
\title{Dynamics near a minimal-mass soliton for a Korteweg--de Vries
  equation}

\author{J.L. Marzuola} \email{marzuola@math.unc.edu}

\author{S. Raynor} \email{raynorsg@wfu.edu}

\author{G. Simpson} \email{gsimpson@umn.edu}

\date{\today}

\begin{abstract}
  We study soliton solutions to a generalized Korteweg - de Vries
  (KdV) equation with a saturated nonlinearity, following the line of
  inquiry of the authors in \cite{MRS1} for the nonlinear
  Schr\"odinger equation (NLS).  KdV with such a nonlinearity is known
  to possess a minimal-mass soliton.  We consider a small perturbation
  of a minimal-mass soliton and identify a system of ODEs, which
  models the behavior of the perturbation for short times.  This
  connects nicely to a work of Comech, Cuccagna \& Pelinovsky
  \cite{Comech:2007p832}.  These ODEs form a simple dynamical system
  with a single unstable hyperbolic fixed point with two possible
  dynamical outcomes. A particular feature of the dynamics are that
  they are non-oscillatory.  This distinguishes the KdV problem from
  the analogous NLS one.
\end{abstract}

\maketitle


\section{Introduction}

We consider a generalized Korteweg-deVries equation equation of the
form
\begin{equation}
  \label{e:skdv}
  u_t + \partial_x \paren{f(u)} + u_{xxx} = 0
\end{equation}
where $f$ is a {\it saturated} nonlinearity; that is, $f$ behaves
subcritically at high intensities and supercritically at low
intensities.  An example of such a nonlinearity is
\begin{equation}
  \label{eq:sat1}
  f(s) = \frac{s^p}{1+\delta s^{p-q}}.
\end{equation}
with $p>5$ and $1<q<5$, and with $\delta >0$ as an additional
parameter.  In the computations we present, we take $p =6$, $q=3$ and
$\delta = \tfrac{1}{4}$.

\begin{figure}
  \includegraphics[width=4in]{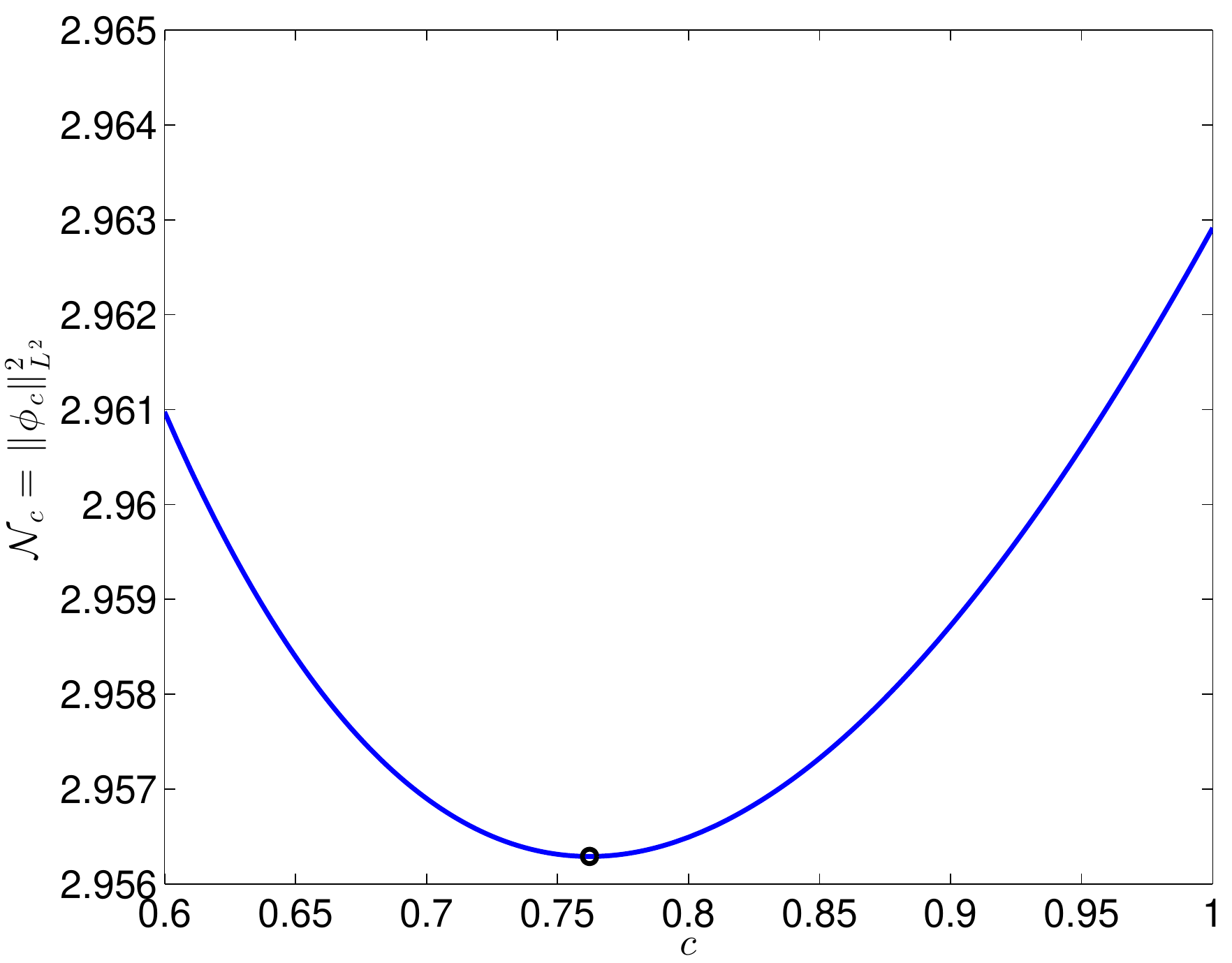}
  \caption{$\calN_c = \norm{\phi_c}^2_{L^2}$ possesses a local minimum
    for the saturated nonlinearity given in \eqref{eq:sat1}. }
  \label{f:solcurves}
\end{figure}

A traveling wave function $u(x,t)=\phi_c(x-ct)$, is a {\em{soliton}}
solution to \eqref{e:skdv} when the profile $\phi_c$ satisfies the ODE
\begin{equation}
  \label{e:soliton}
  - c \phi_c + f(\phi_c) + \partial_{yy}\phi_c=0.
\end{equation}
If we were considering a noncritical power nonlinearity, $f(s) = s^p$,
the equation would admit solitons of arbitrarily small $L^2$-norm.
However, saturated nonlinearities are more restrictive.  For the
nonlinearities we consider, there will be a unique soliton of minimal
$L^2$-norm.  See Figure \ref{f:solcurves} for a plot of the $L^2$-norm
as a function of $c$ for one instance of \eqref{eq:sat1}.  This
critical soliton is denoted $\phi_{c_\star}$.

It is known that there exists an interval $U \subset \R$ so that there
is a soliton solution to \eqref{e:soliton} for each $c \in U$
\cite{GSS}; moreover, $\phi_c(y)$ is a smooth function of $c$ on $U$.
Indeed, via elliptic theory we see that it holds generically that
$\phi \in C^{q+2}$ ; since $\phi > 0$, this then implies that $\phi
\in C^\infty$ for $p,q \in \mathbb{Z}$.  Solitons can be interpreted
as minimizers of the Hamiltonian energy
\begin{equation}
  E(u) = \frac12 \int_\R | \p_x u |^2 - \int_\R F(|u|) dx,
  \label{e:energy}
\end{equation}
where $F(s) = \int_0^s f(t) dt$, subject to the fixed momentum
condition
\begin{eqnarray}
  \label{e:momentum}
  N (u) = \frac12 \int_\R |u|^2 dx,
\end{eqnarray}
with $c$ acting as the Lagrange multiplier.  Equation \eqref{e:skdv}
also conserves the mass:
\begin{eqnarray}
  \label{e:mass}
  I (u) = \int_\R u \ dx.
\end{eqnarray}
By evaluating these conserved quantities at the soliton $\phi_c$, we
get functions of $c$:
\begin{equation}
  \label{e:functionals}
  \calE_c = E (\phi_c), \quad  \calN_c = N(\phi_c) , \quad \calI_c = I (\phi_c).
\end{equation}
The minimal-momentum soliton is found at the value of $c = c_\star$
such that
\begin{equation}
  \label{e:crit_cond}
  \frac{d}{dc} \calN_c = \inner{\phi_c}{\dc \phi_c} = 0.
\end{equation}
This first order condition would also hold at a maximal momentum
soliton.

While the stability of soliton solutions of \eqref{e:skdv} is well
understood away from such critical points, our understanding of the
dynamics near this critical soliton remain incomplete.  Due to the
saturated nature of the nonlinearity, the equation is globally well
posed in $H^2$, so there is no finite-time singularity.  But does the
perturbed soliton converge to some nearby stable state, disperse, or
engage in some other dynamic? It is known \cite{Comech:2007p832}, that
the minimal-mass soliton itself enjoys a purely nonlinear instability.
The purpose of this work is to further examine the dynamics of this
type of solution.

To better understand these dynamics, we consider perturbations of
$\phi_{c_\star}$.  Beginning with the ansatz
\begin{equation}
  u(x,t) = \phi_c(x- ct) + p(x-ct, t)
\end{equation}
for a perturbed soliton, we obtain the following evolution equation
for $p$:
\begin{equation}
  \label{e:expanded}
  p_t = \dy \bracket{-\partial_y^2 p + c p -  f'(\phi_c) p} + \bigo(p^2).
\end{equation}
In order to analyze this equation, we first consider, in Section
\ref{s:lin}, the spectrum of the linearized operator, $A_c$, where
\begin{equation}
  \label{e:linop}
  A_c \equiv \partial_y L_c, \quad L_c \equiv - \partial_{yy} + c - f'(\phi_c).
\end{equation}
An examination of $A_c$ reveals that its generalized kernel has a
dimension of at least two.  At $\cstar$, this dimension increases to
least three, and could be four under special circumstances. Thus,
there will be secular growth of the perturbation at a critical point
generated by the components of the solution parallel to elements of
the generalized kernel.  Even if the perturbation is initially
orthogonal to these unstable directions, the higher order terms will
likely generate unstable contributions.

Such secular growth is eliminated by making a more general ansatz that
allows $c$ to modulate in time. This is given in Section \ref{s:exps},
where we introduce a three dimensional set of scalar parameters,
including the soliton speed and wave center, and allow them to
modulate.  This permits for projection away from the linearly unstable
modes.  We separate the projection of $p$ onto the discrete spectrum
of $A_c$ from its projection onto the continuous spectrum.  We then
discard the continuous projection component while modulating the
remaining parameters to obtain a two dimensional system of first-order
ordinary differential equations.

Finally, in Section \ref{s:nummeth}, we describe the numerical methods
we use to compute key parameters and simulate \eqref{e:skdv}.  Section
\ref{s:results} presents the results of these computations and
discusses their implications.

\section{The Linear Operator}
\label{s:lin}

In this section, we survey the spectral properties of $A_c$ as defined
in \eqref{e:linop}, about $c=c_\star$.  For all values of $c$ that
admit a soliton, one can directly compute
\begin{align}
  A_c(-\partial_y \phi_c) & = 0, \\
  A_c \partial_c \phi_c & = - \partial_y \phi_c.
\end{align}
For the remainder of this article, we will denote differentiation with
respect to $c$ by ${}'$.  The first two elements of the generalized
kernel of $A_c$, then, are:
\begin{equation}
  \label{e:kerA12}
  e_{1,c} = -\partial_y \phi_c, \quad e_{2,c} = \phi_c'.
\end{equation}
At a minimal-momentum soliton (or indeed, any soliton satisfying the
first order condition \eqref{e:crit_cond}), there is a third
independent function $e_{3,\cstar}$ in the generalized kernel of
$A_c$, which satisfies
\begin{equation}
  \label{e:ker3}
  A_{c_\star} e_{3,c_\star} = e_{2,\cstar}.
\end{equation}
To see why such a state exists, consider the adjoint operator,
$A_c^\ast =-L_c \dy$. We immediately compute
\begin{equation}
  A_c^\ast \phi_c  = 0,\quad A_c^\ast D^{-1} \phi_c'  = -\phi_c ,
\end{equation}
where
\begin{equation}
  D^{-1}f \equiv \int_{-\infty}^y f.
\end{equation}
Hence we define:
\begin{equation}
  g_{1,c} = \phi_c, \quad g_{2,c} = D^{-1} \phi_c'.
\end{equation}
Then, by the Fredholm alternative, we see that$A_c f = e_{2,c}$ has a
solution provided
\begin{equation}
  {e_{2,c}}{ g_{1,c}} = \int \phi_c' \phi = \frac{d}{dc}\int
  \frac{1}{2}\phi_c^2 = \calN_c'
\end{equation}
vanishes, which is condition \eqref{e:crit_cond}.  Thus, at a
minimal-mass soliton, there is indeed a third element of the
generalized kernel of $A_\cstar$, which solves
$$A_\cstar e_{3,\cstar}=e_{2,\cstar}.$$
Consequently, there is also a third element in the generalized kernel
of $A_\cstar^\ast$,
\begin{equation}
  \label{e:ker3adj}
  A_\cstar^\ast g_{3,\cstar} = g_{2,\cstar}
\end{equation}
To see if there is a fourth element in the generalized kernel, we
again consider $A_\cstar f = e_{3,\cstar}$ and recognize that we would
need
\begin{equation}
  \begin{split}
    \inner{e_{3,\cstar}}{g_{1,\cstar}} &=
    \inner{e_{3,\cstar}}{A_\cstar^\ast g_{2,\cstar}} = \inner{A_\cstar
      e_{3,\cstar} }{g_{2,\cstar}} \\
    &= \inner{e_{2,\cstar}}{g_{2,\cstar}} = \int \phi_\cstar'
    \paren{\int_{-\infty}^y \phi_\cstar'}dy\\
    & = \int \frac{d}{dy}\frac{1}{2}\paren{\int_{-\infty}^y
      \phi_\cstar' }^2= \frac{1}{2}(\calI_\cstar')^2
  \end{split}
\end{equation}
to vanish. Even at a critical value of $\cstar$ for which
$\calN_\cstar' =0$, it is not generic to observe $\calI_\cstar' = 0$.
Indeed, for the particular nonlinearity $f$ that we consider, our
minimal-momentum soliton will not have this fourth element.  It would
be of interest to find a nonlinearity that does satisfy this
additional degeneracy condition, and to study the dynamics near the
resulting doubly-critical soliton.

A particular challenge, discussed below in Section \ref{s:nummeth}, is
that some of these generalized kernel elements, notably
$e_{3,\cstar}$, $g_{2,\cstar}$ and $g_{3,\cstar}$ are not in $L^2$.
While $e_{3,\cstar}$ vanishes exponentially fast at $+\infty$, it is
only bounded at $-\infty$.  $g_{2,\cstar}$ and $g_{3,\cstar}$ both
vanish at $-\infty$, but at $+\infty$ the former is only bounded and
the latter grows linearly. The reader can find a discussion the
function spaces in which these kernel functions lie in
\cite{Comech:2007p832}.

For later use, we remark that away from $\cstar$, using the implicit
function theorem, there is a scalar $\lambda_c$ and function
$e_{3,c}$, both smooth in $c$, such that
\begin{equation}
  \label{e:lamcont}
  \paren{A_c- \lambda_c I} e_{3,c} = e_{2,c},\quad \lambda_c \equiv -
  \frac{ \calN_c' }{ \langle \phi_c, e_{3,c} \rangle}.
\end{equation}

\section{Modulation Equations}
\label{s:exps}

To overcome the secular growth due to the generalized kernel, we now
permit the equation parameters to modulate about the extremal soliton.
First, we define the moving frame
\begin{equation}
  y(x,t) \equiv  x -\int_0^tc(\sigma)d\sigma - \xi(t).
\end{equation}
Next, we consider a solution $u(x,t)$ which is a perturbed, modulating
soliton,
\begin{equation}
  \label{e:pertmodsol}
  \begin{split}
    u(x,t) &= \phi_{c(t)}\paren{x - \int_0^tc(\sigma)d\sigma - \xi(t)}
    \\
    &\quad+
    p\paren{x-\int_0^tc(\sigma)d\sigma -\xi(t), t}\\
    & = \phi_c(y) + p(y,t).
  \end{split}
\end{equation}
Substituting this into \eqref{e:skdv}, we get
\begin{multline}\label{e:mod}
  p_t + \phi'_c \dot{c} - \partial_y \phi_c\dot{\xi} - p_y \dot{\xi}
  \\= \partial_y L_{c(t)} p -\frac{1}{2}\partial_y \paren{f''(\phi_c)
    p^2} + \text{higher order terms.}
\end{multline}
We let
\begin{equation}
  \label{e:etadef}
  \eta (t)  =  c(t) - \cstar,
\end{equation}
and decompose
\begin{equation}
  \label{e:pdecomp}
  p (x,t)  =  \zeta (t) e_{3,c} + v  = \zeta(t) e_{3,\cstar} +
  \zeta(t) \eta(t) e_{3,\cstar}' + v + \bigo(\zeta\eta^2)
\end{equation}
as in \cite{Comech:2007p832}, equations (3.19--3.24), which results in
\begin{equation*}
  v_t + A_c v = -\dot{\xi} e_{1,c} - (\dot{\eta} - \zeta) e_{2,c} - (\dot{\zeta} - \lambda_c \zeta) e_{3,c} - \dot{\eta} \p_x p + \p_x N.
\end{equation*}
To close the system, we introduce the constraints
\begin{equation*}
  \langle g_{1,c}, v \rangle = \langle g_{2,c} , v \rangle = \langle g_{3,c}, v \rangle = 0.
\end{equation*}

We make several observations about the resulting dynamical system.
The term $\p_x N$ has quadratic and higher-order terms.  We will
preserve only quadratic terms in our computations.  We will also
disregard all coupling to the continuous spectrum, though some of this
may also be of quadratic order.  Thus the quadratic order terms we are
considering in our approximation can be taken to be
$\frac12\zeta^2 \partial_x(f''(\phi_c)e_{3,c}^2)$.  Projecting onto
the canonical spectral functions, the finite dimensional system then
takes the form
\begin{equation}
  \begin{pmatrix}
    \dot{\xi}\\
    \dot{\eta} - \zeta\\
    \dot{\zeta} - \lambda_c \zeta
  \end{pmatrix} = -\zeta^2\mathcal{S}_c(\zeta) ^{-1}\begin{pmatrix}
    \inner{g_{1,c}}{e_{3,c}' - \frac12\p_x(f''(\phi_c)e_{3,c}^2)} \\
    \inner{g_{2,c}}{e_{3,c}' - \frac12\p_x(f''(\phi_c)e_{3,c}^2)}\\
    \inner{g_{3,c}}{e_{3,c}' - \frac12\p_x(f''(\phi_c)e_{3,c}^2)}
  \end{pmatrix},
\end{equation}
Under this approximation, $\xi$ is slaved to $\eta$ and $\zeta$.
There is only weak coupling between $\xi$ and the other parameters
through $v$.

Continuing with the above assumptions,
\begin{equation}
  \mathcal{S}_c(\zeta) = \mathcal{T}_c
  + \zeta \begin{pmatrix}
    - \inner{g_{1,c}}{\dx e_{3,c}} & \inner{g_{1,c}}{e_{3,c}'} & 0\\
    - \inner{g_{2,c}}{\dx e_{3,c}} & \inner{g_{2,c}}{e_{3,c}'} & 0\\
    - \inner{g_{3,c}}{\dx e_{3,c}} & \inner{g_{3,c}}{e_{3,c}'} & 0
  \end{pmatrix} = \mathcal{T}_c + \zeta \hat{\mathcal{S}}_c,
\end{equation}
where $(\mathcal{T}_c)_{jk} = \langle g_{j,c}, e_{k,c} \rangle$.
Assuming $\mathcal{T}_c$ and $\mathcal{S}_c$ are $\bigo(1)$ and
$\zeta$ is sufficiently small, we can make the approximation
$\mathcal{S}_c\approx \mathcal{T}_c$, and thus obtain, for appropriate
vectors $R_c$ and $Q_c$:
\begin{align*}
  \begin{pmatrix}
    \dot{\xi}\\
    \dot{\eta} - \zeta\\
    \dot{\zeta} - \lambda_c \zeta
  \end{pmatrix} & = -\zeta^2\mathcal{S}_c( \zeta)^{-1}\begin{pmatrix}
    \inner{g_{1,c}}{e_{3,c}' -\frac{1}{2} \p_x(f''(\phi_c)e_{3,c}^2)} \\
    \inner{g_{2,c}}{e_{3,c}' -\frac{1}{2} \p_x(f''(\phi_c)e_{3,c}^2)}\\
    \inner{g_{3,c}}{e_{3,c}' -\frac{1}{2} \p_x(f''(\phi_c)e_{3,c}^2)}
  \end{pmatrix}\\ & = -\zeta^2 \mathcal{T}_{c}^{-1} \begin{pmatrix}
    \inner{g_{1,c}}{e_{3,c}' -\frac{1}{2} \p_x(f''(\phi_c)e_{3,c}^2)} \\
    \inner{g_{2,c}}{e_{3,c}' -\frac{1}{2} \p_x(f''(\phi_c)e_{3,c}^2)}\\
    \inner{g_{3,c}}{e_{3,c}' -\frac{1}{2} \p_x(f''(\phi_c)e_{3,c}^2)}
  \end{pmatrix} + \bigo(\zeta^3)  \\
  & = -\zeta^2 \mathcal{T}_c^{-1} R_c + \bigo(\zeta^3) \\
  & = -\zeta^2 Q_c + \bigo(\zeta^3)
\end{align*}

Therefore, the leading order equations, subject to these
approximations, are
\begin{subequations}
  \begin{align}
    \label{e:eq3} \dot \xi & = - Q_{1,c} \zeta^2, \\
    \label{e:eq1} \dot{\eta} - \zeta & = -Q_{2,c}\zeta^2,  \\
    \label{e:eq2} \dot{\zeta} - \lambda_c \zeta & = -Q_{3,c}\zeta^2.
  \end{align}
\end{subequations}
Making the Taylor expansions of $\lambda_c$ the $Q_{j,c}$'s about
$c=\cstar$, and omitting the $\xi$ equation, we obtain the
quadratically nonlinear ODE system
\begin{subequations}\label{e:simplified_ccp}
  \begin{align}
    \dot{\eta} - \zeta & = -Q_{2,\cstar}\zeta^2\\
    \dot{\zeta} - \lambda'_{\cstar} \eta\zeta & =
    -Q_{3,\cstar}\zeta^2.
  \end{align}
\end{subequations}
Critical points of the system are found at $\zeta =0$, $\eta=\eta_0$
where $\eta_0$ is arbitrary, and at $\zeta = \frac{1}{Q_2}$, $\eta
=\frac{Q_3}{\lambda_\cstar' Q_2}$. The latter isolated critical point
is a saddle point in the first quadrant.

\begin{figure}
  \includegraphics[width=4in]{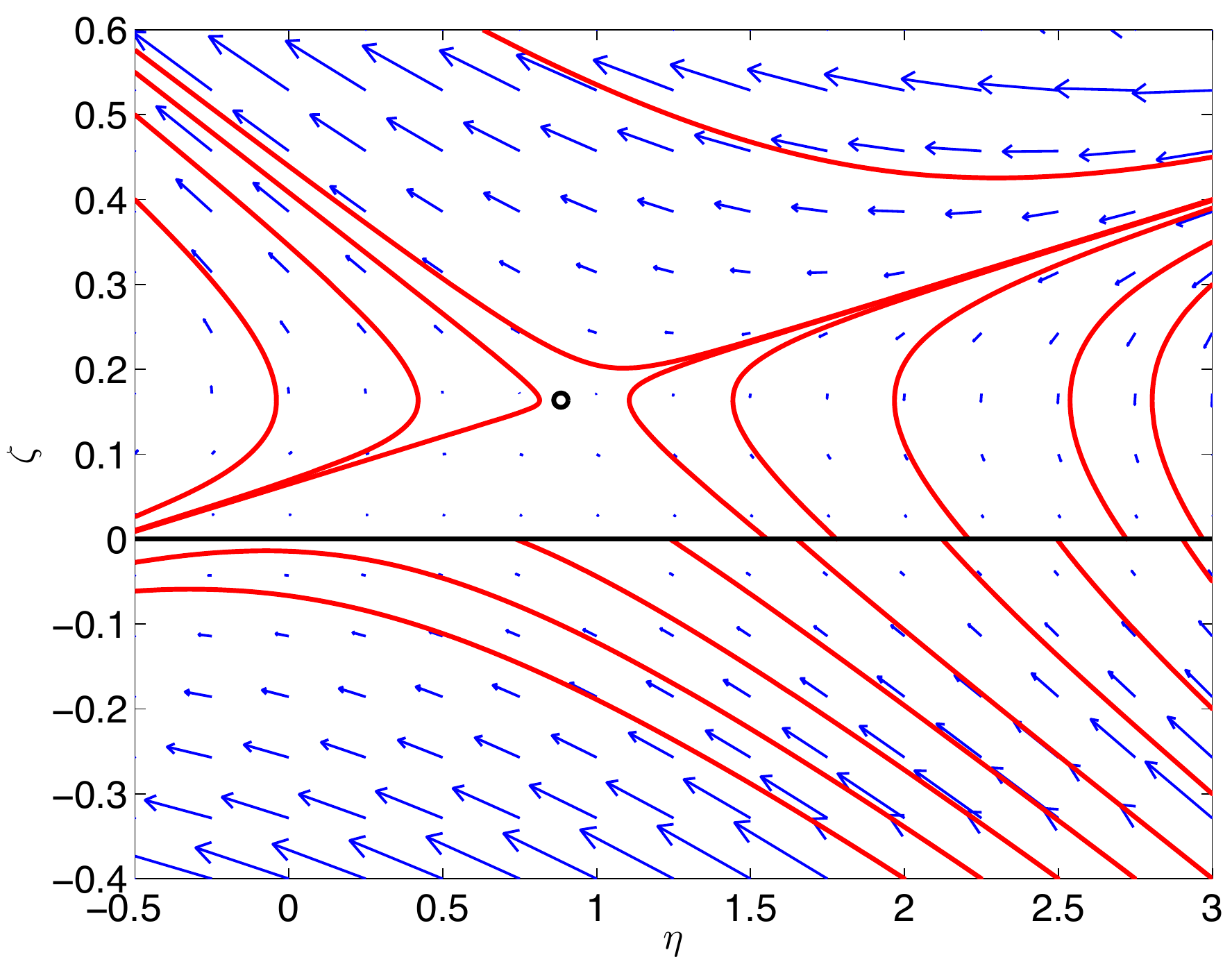}
  \caption{A plot of a representative phase plane diagram.  The
    hyperbolic fixed point is indicated by the $\circ$ while the
    nonisolated critical points are indicated by the solid line $\zeta
    =0$.}
  \label{f:pplane}
\end{figure}

For the $\zeta =0$ critical points the linearized problem is
\[
\begin{pmatrix}
  0 & 1\\
  0 & \lambda'_\cstar \eta_0
\end{pmatrix}
\]
The eigenvalues are $\lambda'_\cstar \eta_0$ and $0$.  Thus, depending
on the sign of $\lambda'_\cstar \eta_0$, the solution is either
linearly stable or unstable.  It is worth noting in this context that
$\lambda_c = \frac{-{\mathcal{N}}_c'}{\inner{\phi_c}{e_{3,c}}}$.
Thus,
$\lambda_\cstar'=\frac{-{\mathcal{N}_\cstar}''}{\inner{\phi_c}{e_{3,c}}}$
since ${\mathcal{N}}'_\cstar=0$.  Also note that, according to
\cite{Comech:2007p832}, near $\cstar$ we have that
${\inner{\phi_c}{e_{3,c}}} > 0$.  Thus, we expect that the sign of
$\lambda'_\cstar$ depends solely on whether we are at a minimal or
maximal soliton.  In each case, we expect to see that, depending on
the sign of $\eta_0$, the critical point at $(0,\eta_0)$ is either
linearly stable or linearly unstable. There is semi-stability at
$(0,0)$, depending on the sign of the initial perturbation $\eta_0$.
See Figure \ref{f:pplane} for a representative saturated KdV phase
plane diagram.

\section{Computational Methods}\label{s:nummeth}

In this section we briefly outline the computations needed to make a
comparison between \eqref{e:simplified_ccp} and the KdV equation,
\eqref{e:skdv}.  Motivated by \eqref{e:pertmodsol}, \eqref{e:etadef}
and \eqref{e:pdecomp}, we will take initial conditions of the form
\begin{equation}
  \label{e:initdata}
  u_0 = \phi_{\cstar} ( x) + \eta_0 e_{2,\cstar} + \zeta_0 e_{3,\cstar}.
\end{equation}

\subsection{Spectral Computations}
\label{s:speccomp}

To compute the coefficients appearing in \eqref{e:simplified_ccp}, the
$Q_j$'s and $\lambda_\cstar'$, we must compute
\begin{itemize}
\item The generalized kernels of $A_\cstar$ and $A_\cstar^\ast$ from
  which we can get the inner products
  $\inner{g_{j,\cstar}}{e_{k,\cstar}}$ which the matrix
  $\mathcal{T}_\cstar$ comprises;
\item $\dx e_{3,\cstar}$ and $e_{3,\cstar}'$ to obtain the $R_\cstar$
  vector, which, with $\mathcal{T}_\cstar$, allows us to obtain the
  $Q_\cstar$ vector;
\item $\lambda_\cstar'$, which can be obtained by differentiating
  \eqref{e:lamcont} and then computing $\phi_\cstar''$.
\end{itemize}

The first few elements of the generalized kernel, $e_{1,\cstar} = \dx
\phi_\cstar$, $e_{2,\cstar} = \phi_\cstar'$ and $g_{1,\cstar} =
\phi_\cstar$ are readily obtained using the the $\sinc$ discretization
method previously used by the authors in \cite{MRS1}.  Briefly, this
approach solves equations like
\begin{equation}
  L_c f = g
\end{equation}
using a $\sinc$ discretization of $f$, $g$ and $L_c$, provided $g$ is
localized.  Derivatives of functions are easy to obtain using the
discretized differentiation matrices, and $L^2$ inner products are
just finite dimensional inner products, multiplied by the uniform grid
spacing.  To obtain the minimal-momentum soliton, we use a
root-finding algorithm to solve $\calN_c' = 0$. See
\cite{lund1992smq,stenger1993nmb,weideman2000mdm,SimpsonSpiegelman}
for additional details on the $\sinc$ discretization method.

Computing the other elements requires a bit more care as they are not
$L^2$-localized.  First, let
\begin{equation}
  \Theta(y)   \equiv \int_{+\infty}^y e_{2,\cstar}.
\end{equation}
Then $e_{3,\cstar}$ solves
\begin{equation}
  \label{e:e3Theta}
  L_\cstar e_{3,\cstar} = \Theta, \quad \lim_{y\to +\infty} e_{3,\cstar}(y) = 0.
\end{equation}
Given $\sinc$ discretization of $\phi_\cstar'$, we can readily
integrate to obtain $\Theta$ using the techniques given in
\cite{Stenger:1981ws}.  The use of such a quadrature tool was not
required in \cite{MRS1} as all of the NLS kernel elements belonged to
$L^2$.  This also gives us $g_{2,\cstar}$.

Assuming that $e_{3,\cstar}$ grows, at most, algebraically at
$-\infty$ we can drop the $f'(\phi_\cstar)e_{3,\cstar}$ at large
negative values of $y$ term to estimate
\begin{equation}
  -\dx^2 e_{3,\cstar} + \cstar e_{3,\cstar} \approx -\calI_\cstar'\neq 0.
\end{equation}
This approximation implies that $e_{3,\cstar}$ is actually bounded at
$-\infty$:
\begin{equation}
  \lim_{y\to -\infty} e_{3,\cstar} = -\frac{1}{\cstar}\calI_\cstar'.
\end{equation}
One method of computing $e_{3,\cstar}$ is to split it into a piece
which has the above asymptotics, and a spatially localized piece,
\begin{equation}
  e_3\equiv e_{3}^{(1)} + e_{3}^{(2)},
\end{equation}
where $e_{3}^{(1)}\in L^2$ and $e_{3}^{(2)}$ vanishes at $+\infty$.
Using the above estimate, we set
\begin{equation}
  \begin{split}
    \label{e:e32}
    e_{3,\cstar}^{(2)}(x) \equiv e_{3,\cstar}^{-} &\cdot \chi_-(x) \\
    = -\frac{1}{\cstar}\calI_\cstar'&\cdot \frac{1}{2} \paren{1 +
      \tanh(-x)}.
  \end{split}
\end{equation}
We have some flexibility in selecting $\chi_-$.  The essential feature
is that it should not contribute anything at $+\infty$, while
capturing the known asymptotic behavior at $-\infty$.  We then solve
\begin{equation}
  L_\cstar e^{(1)}_{3,\cstar} = \Theta - L_\cstar e_{3,\cstar}^{(2)}.
\end{equation}
As the righthand side is now localized at both $\pm\infty$, we obtain
$e^{(1)}_{3,\cstar}$.

The adjoint problem is similar, but requires slightly more care.
First, we solve
\begin{equation}
  L_\cstar h_{3,\cstar} = g_{2,\cstar}
\end{equation}
and then integrate to obtain $g_{3,\cstar}$.  While $e_{3,\cstar}$ was
asypmtotically constant, $g_{3,\cstar}$ will have linear growth at
$+\infty$.  The other function which requires such an asymptotic
splitting is $e_{3,\cstar}'$,

To compute $\lambda_\cstar'$, we compute
\begin{equation}
  \lambda_\cstar' =
  -\frac{\calN_\cstar''}{\inner{\phi_\cstar}{e_{3,\cstar}}} = -
  \frac{\inner{\phi_\cstar}{\phi_\cstar''}+ \inner{\phi_\cstar'}{\phi_\cstar'}}{\inner{\phi_\cstar}{e_{3,\cstar}}};
\end{equation}
${\phi_\cstar''}$ is $L^2$-localized and obtained by solving
\[
L_\cstar {\phi_\cstar''} = - 2 \phi_\cstar' + f''(\phi_\cstar)
(\phi_\cstar' )^2.
\]
Computing the various inner products, we obtain the matrix
$\mathcal{T}_\cstar$ and the vector $R_\cstar$, from which we can
solve for $Q_\cstar$.  This provides us with all coefficients in the
ODE system.


%

\subsection{A Finite Difference Method for KdV Type Equations}

Integrating \eqref{e:skdv} with initial conditions of the form
\begin{equation}
  \phi_{\cstar +\eta_0} + \zeta_0 e_{3,\cstar}
\end{equation}
requires some care, as $e_{3,\cstar}$ is not spatially localized.
However, since it is asymptotically constant, we can, to leading
order, use approximate Neumann boundary conditions at $\pm x_{\max}$,
the edges of our computational domain:
\begin{equation}
  \label{e:numericalbcs}
  \partial_x u=\partial_x^2 u = 0, \quad \text{at $x=\pm x_{\max}$}.
\end{equation}
Using this approximation, we can then solve \eqref{e:skdv} using a
linearized implicit method formulated in \cite{Djidjeli:1995p13052}.
This method has second order accuracy in time, with spatial accuracy
given by the quality of our finite difference approximations of the
derivatives.  In this work, we use second order symmetric estimates of
the first and third derivatives.

\subsection{Extrapolating and Matching Discretizations}

A challenge in using our $\sinc$ approximations of the kernel
functions is that they are given on one discretized mesh which may not
be sufficiently large to employ the approximate boundary conditions
\eqref{e:numericalbcs} for our time dependent simulation.  To overcome
this, we use the farfield asymptotics of these elements to extrapolate
onto a larger domain with a given mesh spacing.

\begin{figure}
  \includegraphics[width=4in]{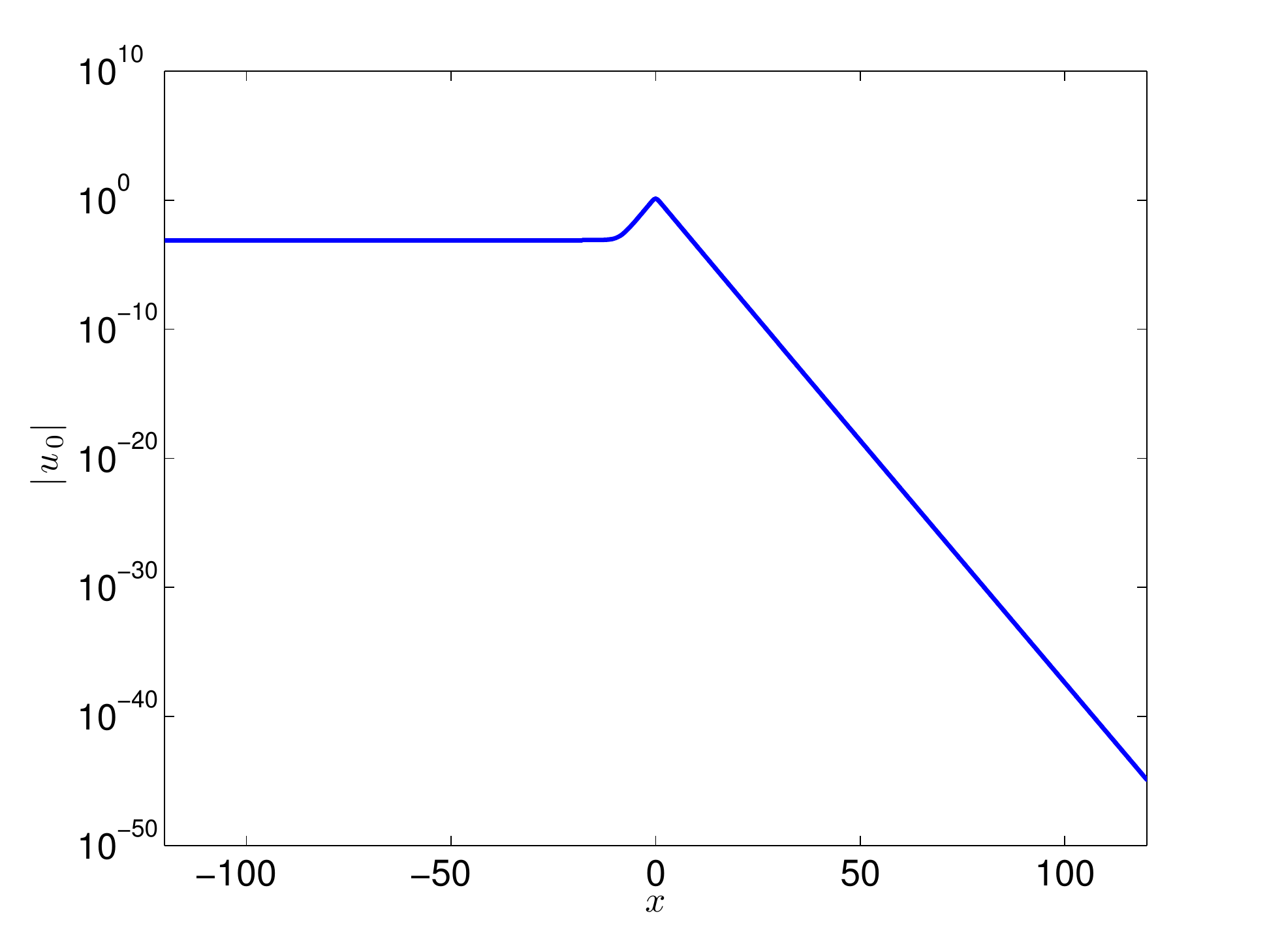}
  \caption{A log plot of the extended initial data.}
  \label{f:logplot}
\end{figure}

\begin{figure}
  \includegraphics[width=4in]{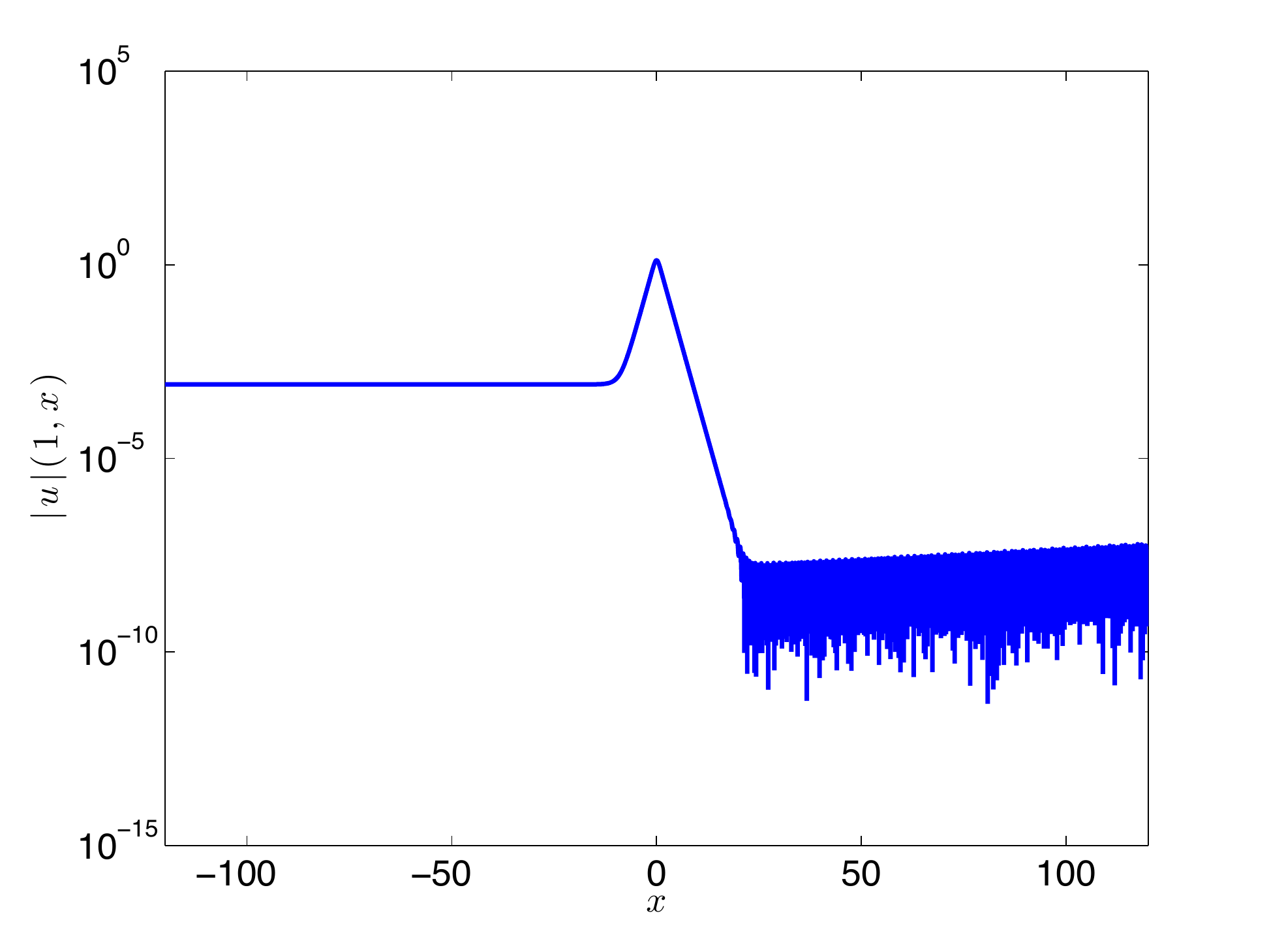}
  \caption{A semilog plot of the numerical solution to our PDE at $t =
    1$ showing the oscillatory back scattering from the boundary at $x
    = R$.}
  \label{f:pdeosc}
\end{figure}

Numerically, we discretize on a short interval, $[-R_{\rm sol},R_{\rm
  sol}]$ to compute the soliton using the iterative $\sinc$ method
from \cite{MRS1}.  We then asymptotically extend $u_0$ to a much
larger interval, $[-R_{\rm as},R_{\rm as}]$ for $R_{as}$ large
relative to where we desire to have the boundary.  In particular, we
extend using the asymptotics
\begin{eqnarray*}
  \phi_{\cstar} (x) & \sim & \alpha_1 e^{-\sqrt{\cstar} |x| }, \ \ \text{as } x \to \pm \infty , \\
  e_{2,\cstar}  (x) & \sim & \alpha_2 x e^{-\sqrt{\cstar} |x| }, \ \ \text{as } x \to \pm \infty ,   \\
  e_{3,\cstar} (x) & \sim &  \alpha_3 x^2 e^{-\sqrt{\cstar} |x| }, \ \ \text{as } x \to  \infty ,  \\
  e_{3,\cstar} (x) & \sim & \tfrac{1}{c} \partial_c \mathcal{I}' -  \alpha_4 x^2 e^{-\sqrt{\cstar} |x| }, \ \ \text{as } x \to  -\infty. \\
\end{eqnarray*}
The asymptotics of $\phi$ are standard, those of $e_{2,\cstar}$ arise
from the commutator relation $[\Delta, x] f = 2 \partial_x f$, and
those for $e_{3,\cstar}$ arise from integration by parts, given the
asymptotics of $e_{2,\cstar}$.  To observe that such a continuation is
nicely continuous and avoid boundary effects from the iterative
methods, we actually choose to extend $\phi$, $e_{2,\cstar}$ and
$e_{3,\cstar}$ from values determined of distance $1$ from $\pm
R_{sol}$.  We include a log plot of an initial condition with
$\eta_0$, $\zeta_0> 0$ in Figure \ref{f:logplot}.  We then a linear
interpolation to have an evenly spaced grid on another still large
interval, $[-R_{\rm pde},R_{\rm pde}]$ with $R_{\rm pde} < R_{\rm
  as}$, on which the simulation is perofrmed.  Here, $R_{\rm pde}$ is
chosen large enough to minimize boundary interactions during the
numerical integration, which in KdV simulations appear quite quickly
due to the dispersion relation; see Figure \ref{f:pdeosc}.

To summarize we have selected three different domains, $R_{\rm sol}$ ,
$R_{\rm as}$ and $R_{\rm pde}$, on which we respectively compute the
soliton, match asymptotics at $\pm \infty$, and then linearly
interpolate onto a uniform grid to solve the PDE, with $R_{\rm sol} <
R_{\rm pde} < R_{\rm as}$.The simulation then proceeds with a linearly
implicit finite difference scheme using a split step time
discretization. In our results, our PDE simulation contains many
oscillations that diminish as the boundary effects are minimized.

\subsection{Extracting Parameters}
\label{s:extract}

As we aim to compare the ODE system \eqref{e:simplified_ccp} with the
PDE, we will need to find a way to extract $\xi$, $\eta =c - \cstar$,
and $\zeta$ from $u(x,t)$,
\begin{equation}
  u(x,t) = \phi_{c}(x - \smallint c - \xi) + \zeta e_{3,c}(x - \smallint c
  - \xi)  + v(x - \smallint c - \xi).
\end{equation}
We estimate the wave speed by computing the center of mass of $u$ at
each time step, and estimating its speed by finite differences; this
gives us $c = \cstar + \eta$. Unfortunately, there is some ambiguity
between the rate at which the wave moves due to the speed, $c$, and
changes in the phase, $\dot\xi$; we are only able to estimate,
collectively,
\begin{equation}
  \partial_t y(x,t) = -c - \dot \xi,
\end{equation}
and assume that this is dominated by the $c$, at least for the time
scales we study.  Indeed, on the time scales over which we numerically
integrate, the $\dot \xi$ term is quadratic in $\zeta$, which must
remain small for our computations to remain accurate.  We then
integrate the wave speed by quadrature to estimate the shift.

Next, we estimate $\zeta$ by projecting onto $g_1$;
\begin{equation}
  \label{e:zeta}
  \zeta(t) = \frac{\langle u(\cdot +, t), g_{1,c} \rangle - \frac{1}{2} c(t)^2 \langle e_{2,c}',
    g_{1,c} \rangle}{\langle e_{3,c}, g_{1,c} \rangle + \cstar \langle e_{3,c}', g_{1,c}
    \rangle },
\end{equation}
which is done by assuming that on the time scales we consider
$g_{1,c}$, $e_{2,c}'$, $e_{3,c}$ and $e_{3,c}'$ are well-approximated
by their values at $c = \cstar$.

\section{Shadowing Results}
\label{s:results}
We now take as an initial condition \eqref{e:initdata} and study the
evolution for different $\eta_0$ and $\zeta_0$.

\begin{figure}
  \subfigure{\includegraphics[width=2.46in]{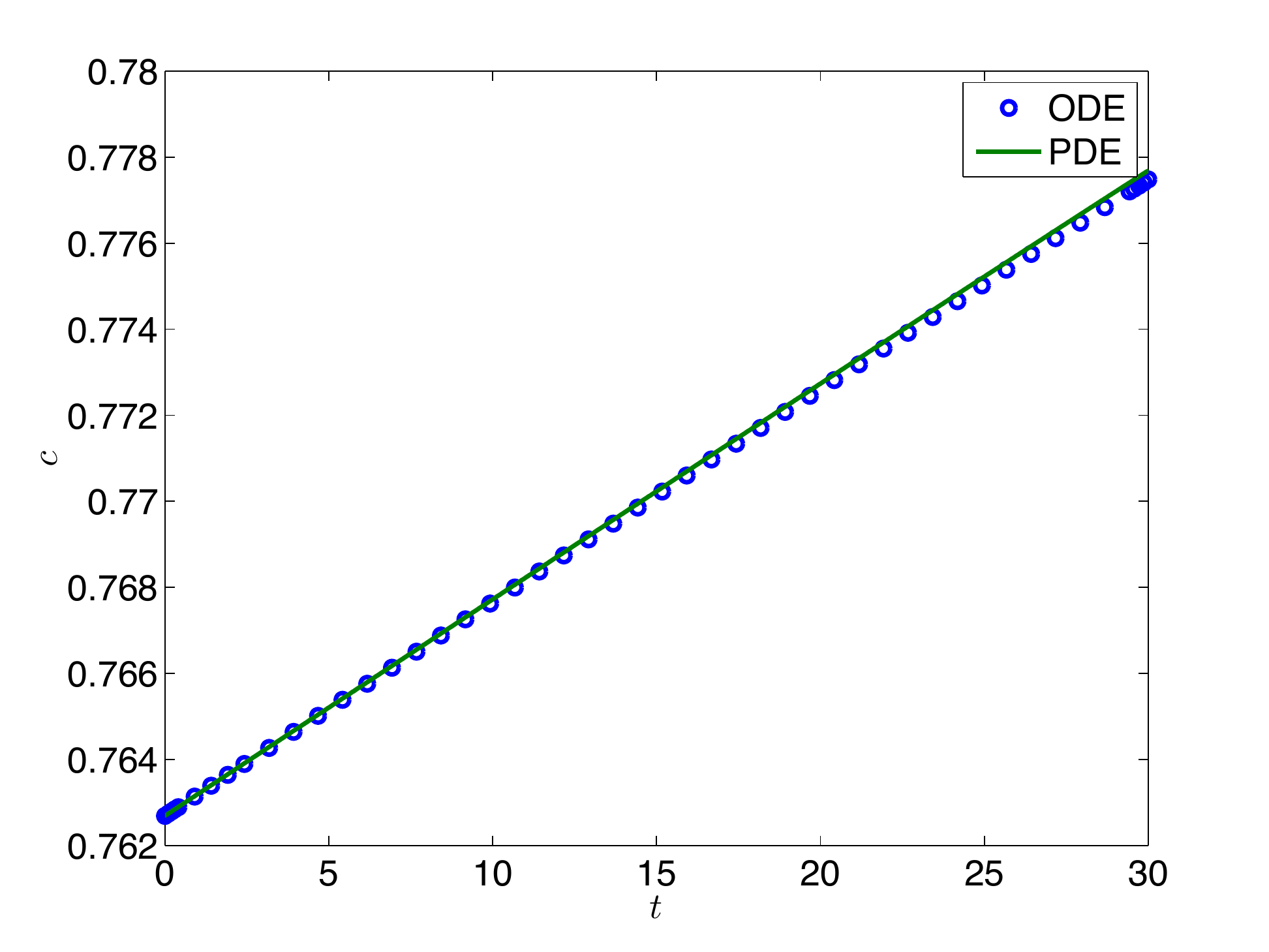}}
  \subfigure{\includegraphics[width=2.46in]{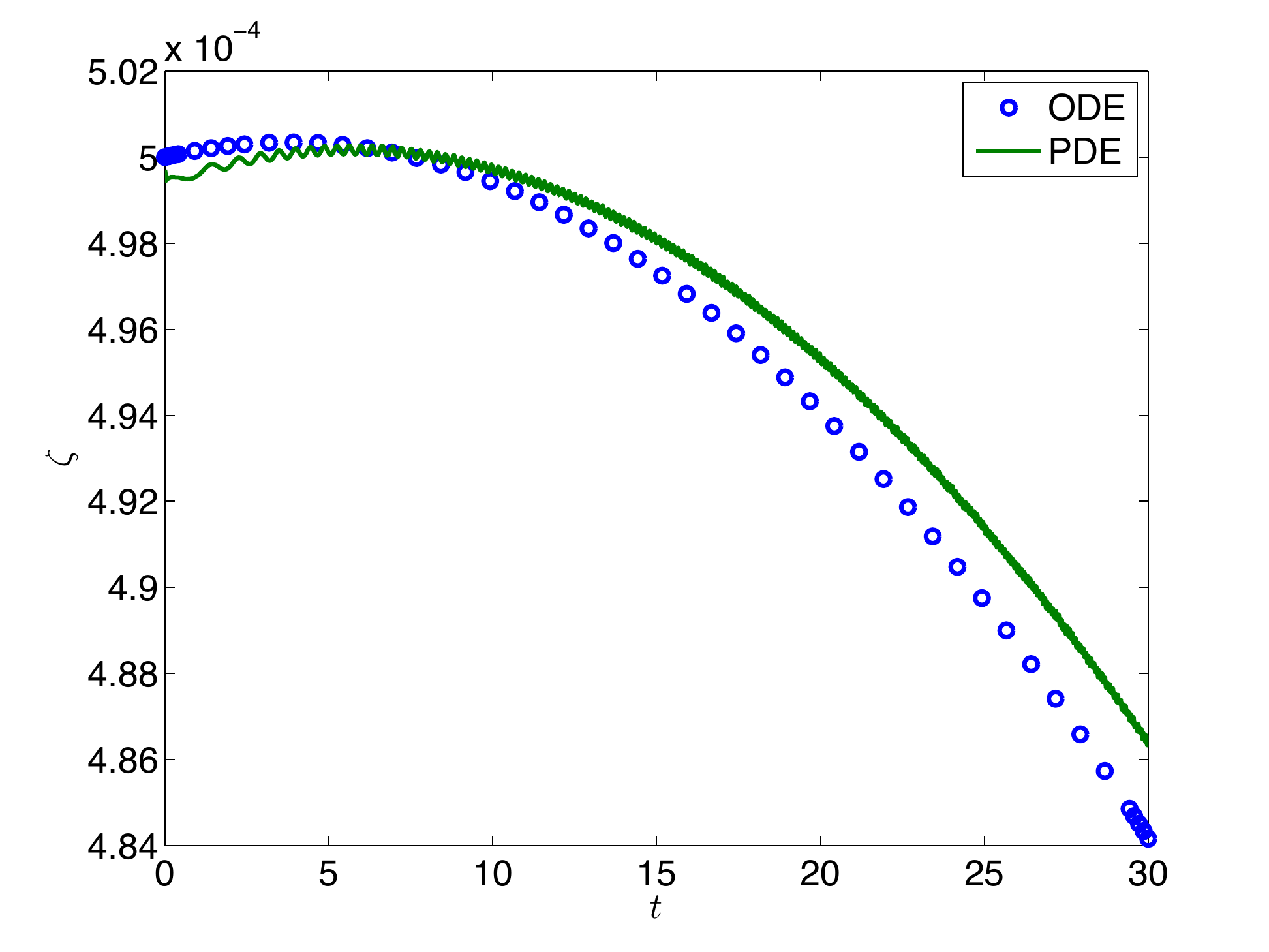}}
  \caption{A comparison between the PDE and the ODE with when $\eta_0
    = 5\times 10^{-4}$ and $\zeta_0 = 5\times 10^{-4}$.}
  \label{f:cmod++}
\end{figure}

\begin{figure}
  \subfigure{\includegraphics[width=2.46in]{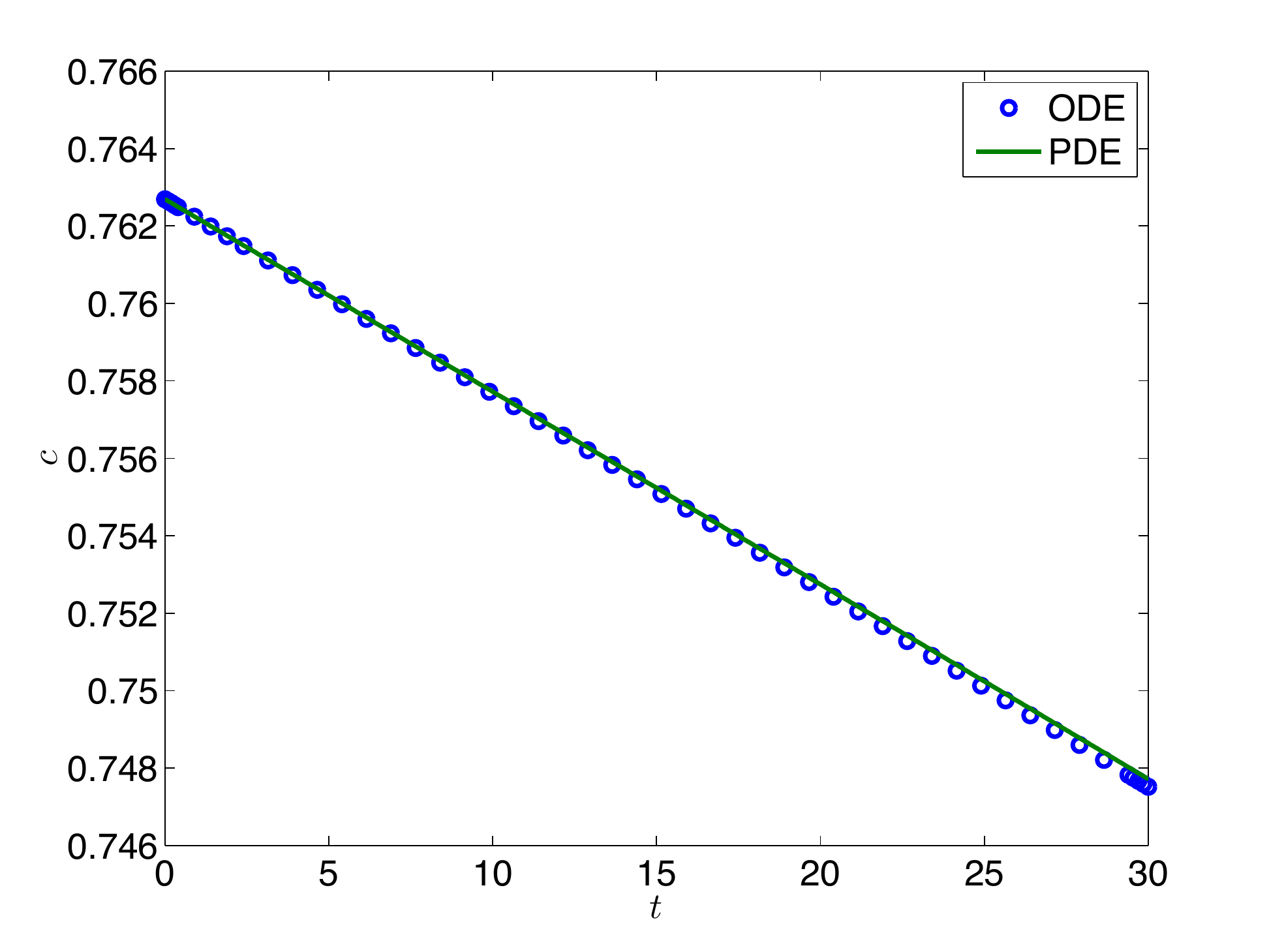}}
  \subfigure{\includegraphics[width=2.46in]{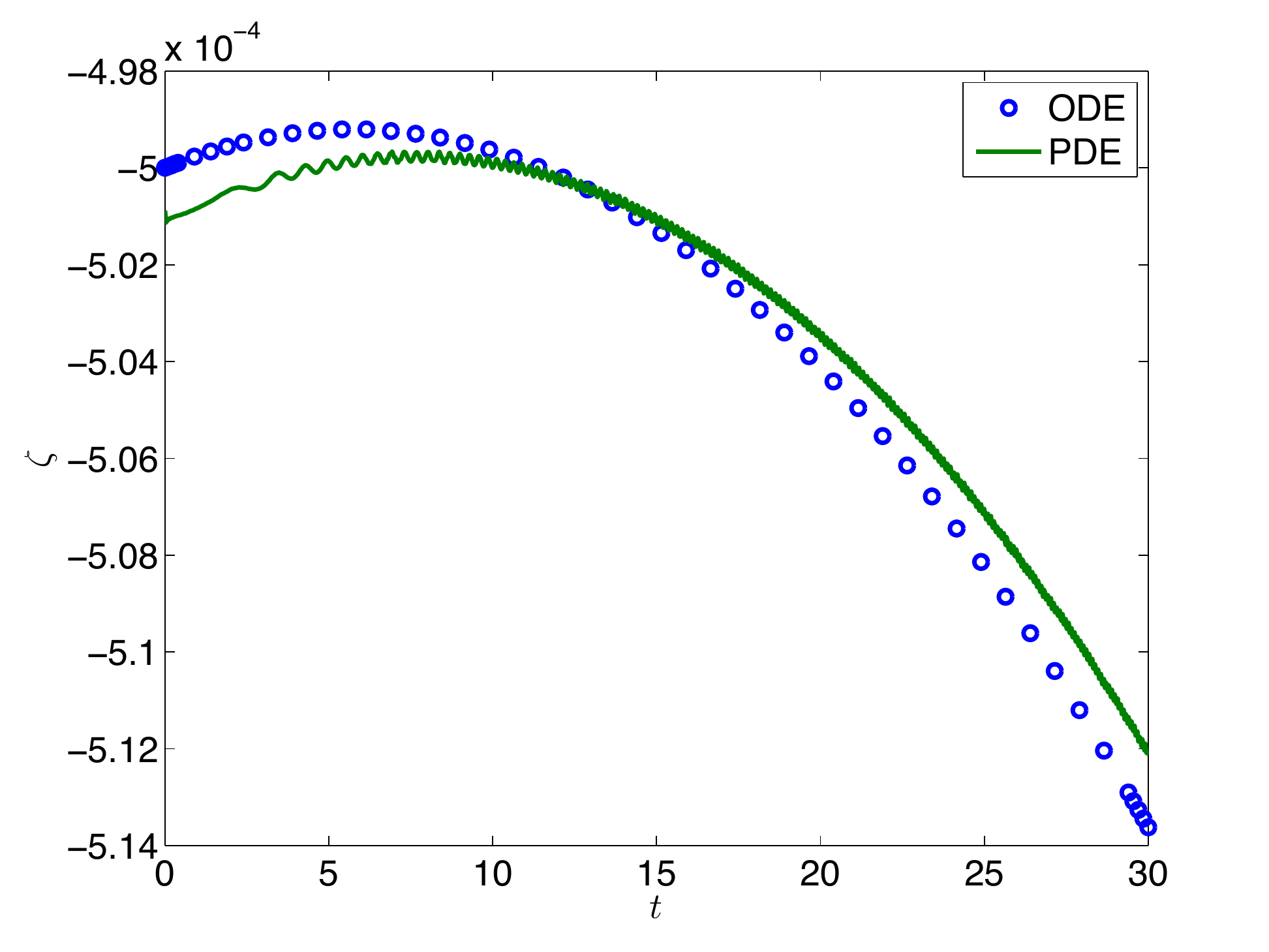}}
  \caption{A comparison between the PDE and the ODE with when $\eta_0
    = 5\times 10^{-4}$ and $\zeta_0 = -5\times 10^{-4}$.}
  \label{f:cmod+-}
\end{figure}

\begin{figure}
  \subfigure{\includegraphics[width=2.46in]{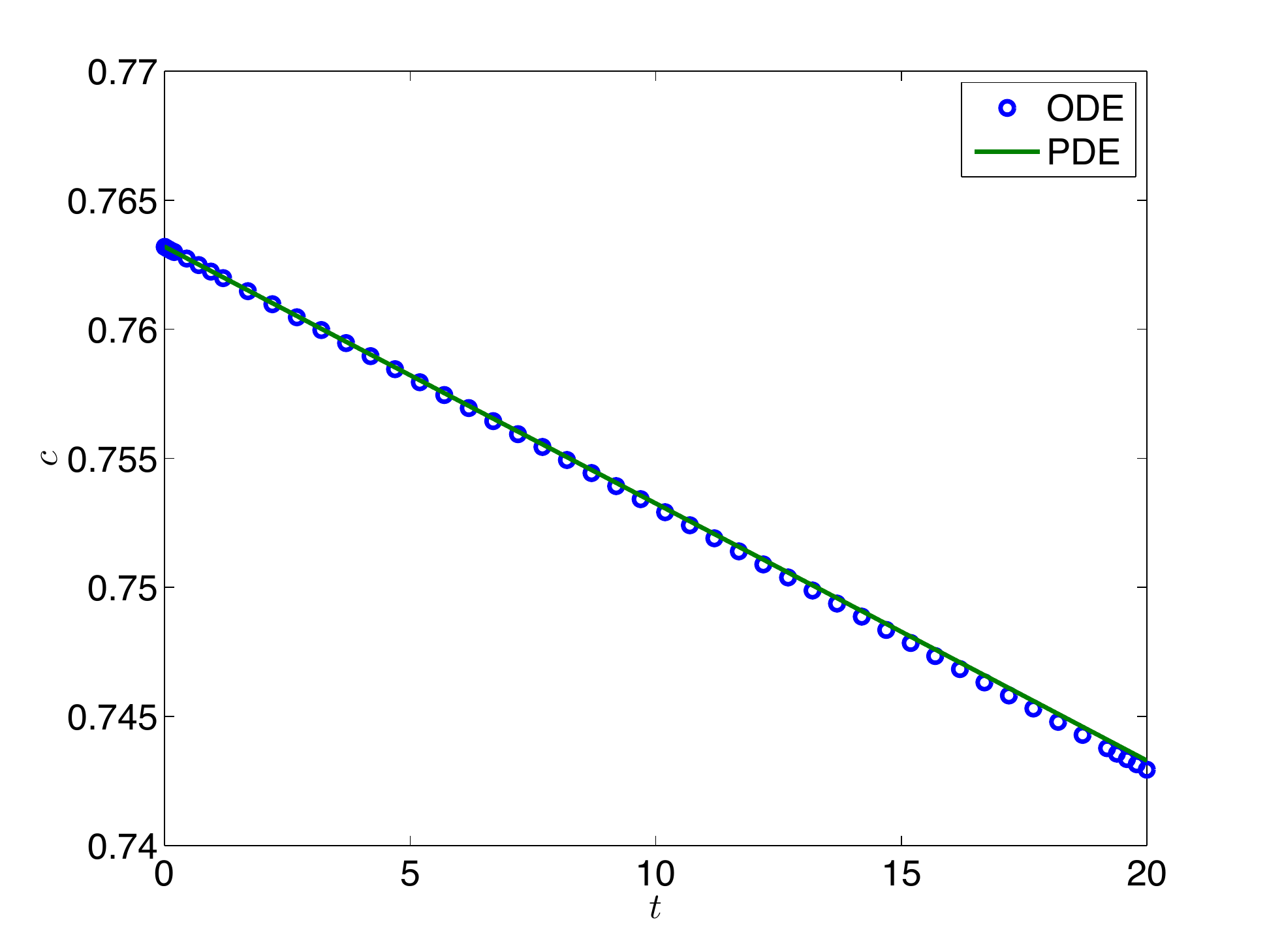}}
  \subfigure{\includegraphics[width=2.46in]{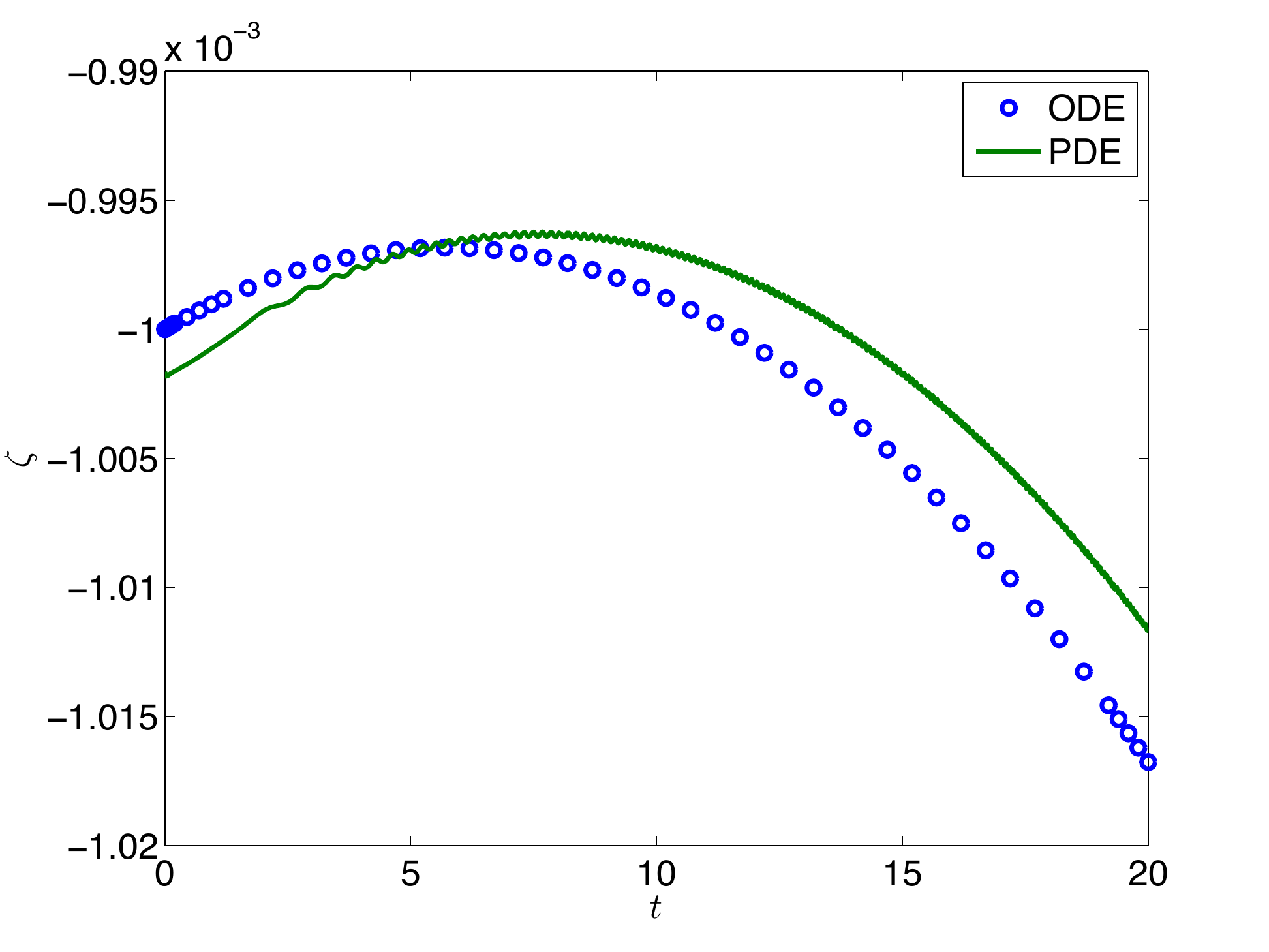}}
  \caption{A comparison between the PDE and the ODE with $\eta_0 =
    10^{-3}$ and $\zeta_0 = -10^{-3}$.}
  \label{f:cmod1}
\end{figure}

The results appear as Figures \ref{f:cmod++} and \ref{f:cmod+-}, in
which we take initial data that begins in the first and third quadrant
of the phase plane \ref{f:pplane} and compare the projection of our
integrated numerical PDE to the predicted ODE dynamics with domain
size $R_{pde} = 120.0$, $N = 10^6$ spatial grid points, time of
integration $T = 30.0$, and time step $h_t = 10^{-4}$.  The remaining
cases display rather similar behaviors.  In \ref{f:cmod1}, we observe
that by taking larger initial $\eta_0 > 0$, $\zeta_0 <0$, our
solutions diverge from the predictive dynamics on a shorter time scale
($T = 20.0$) with otherwise comparable parameters as above.  Since the
nature of KdV is to move to the right, in order to lessen boundary
interaction, we solve the PDE in a moving reference frame around base
velocity $\cstar$ and implement the schemes to project onto the ODE
parameters $c$ and $\zeta$ as in Section \ref{s:extract}.

Implementing these approximative schemes and comparing the evolution
of the corresponding ODE in \eqref{e:simplified_ccp}, the figures show
that for long times the dynamics indeed fit the predicted dynamics.
For small enough perturbations of the minimal-mass soliton, the
dynamical system predicts that the orbits travel very slowly towards
the stable or unstable manifolds.  Hence, we only follow the orbits on
time scales where the parameter $\zeta$ has made a large motion in its
orbit.  The $c$ component varies essentially linearly on this scale
however.  As a result, we observe that \eqref{e:simplified_ccp} is a
good model for perturbations close to the minimal mass.  As in the NLS
case, there is a perturbation of the minimal mass solution leading to
dynamics that move $c$ to smaller values where the corresponding
solitons are linearly unstable.  However, continuing along this
trajectory is forbidden by mass conservation as seen in Figure
\ref{f:solcurves}.  We postulate, as we did for the corresponding
Schr\"odinger dynamics in \cite{MRS1}, that this could be an energy
transfer mechanism to the continuous spectrum in the
infinite dimensional system.  This would eventually lead to
dispersion.  However, as we are here working with perturbations that
are not in $L^2$, it is not possible to compare to known
dispersive solutions as was done for Schr\"odinger.


\appendix
\section{Details of Numerical Methods}

Using the sinc methods described in Section \ref{s:speccomp} and
similarly applied in \cite{MRS1, SimpsonSpiegelman}, we compute the
parameters for system \eqref{e:simplified_ccp}.  The convergence of
these parameters, as a function of the number of grid points, is given
in the Table \ref{t:conv}.

\begin{table}[ht]
  \caption{Convergence of the ODE system parameters as a function of the
    number of grid points, $M$.}
  \label{t:conv}
  \begin{tabular}{r l l l l l}
    \hline
    \hline
    $M$ & $\cstar$ &  $\lambda_c'$ & $Q_1$ & $Q_2$ & $Q_3$\\
    \hline
    20 & 0.76419938 & -0.18921573 & 4.90659351 & 4.14952911 & -1.21924052\\ 
    40 & 0.76214845 & -0.19352933 & 3.33351916 & 5.02642785 & -1.35439215\\ 
    60 & 0.76218663 & -0.19276719 & 2.51228473 & 5.22964908 & -1.36903497\\ 
    80 & 0.76218815 & -0.19262966 & 2.21019564 & 5.28041633 & -1.37169953\\ 
    100 & 0.76218822 & -0.19260110 & 2.10124978 & 5.29465657 & -1.37230370\\ 
    200 & 0.76218823 & -0.19259143 & 2.03532641 & 5.30133115 & -1.37253316\\ 
    300 & 0.76218823 & -0.19259139 & 2.03442654 & 5.30138737 & -1.37253443\\ 
    400 & 0.76218823 & -0.19259139 & 2.03440251 & 5.30138854 & -1.37253445\\ 
    500 & 0.76218823 & -0.19259139 & 2.03440202 & 5.30138858 & -1.37253445\\ 
    \hline
  \end{tabular}
\end{table}

\begin{figure}[ht]
  \subfigure{\includegraphics[width=2.46in]{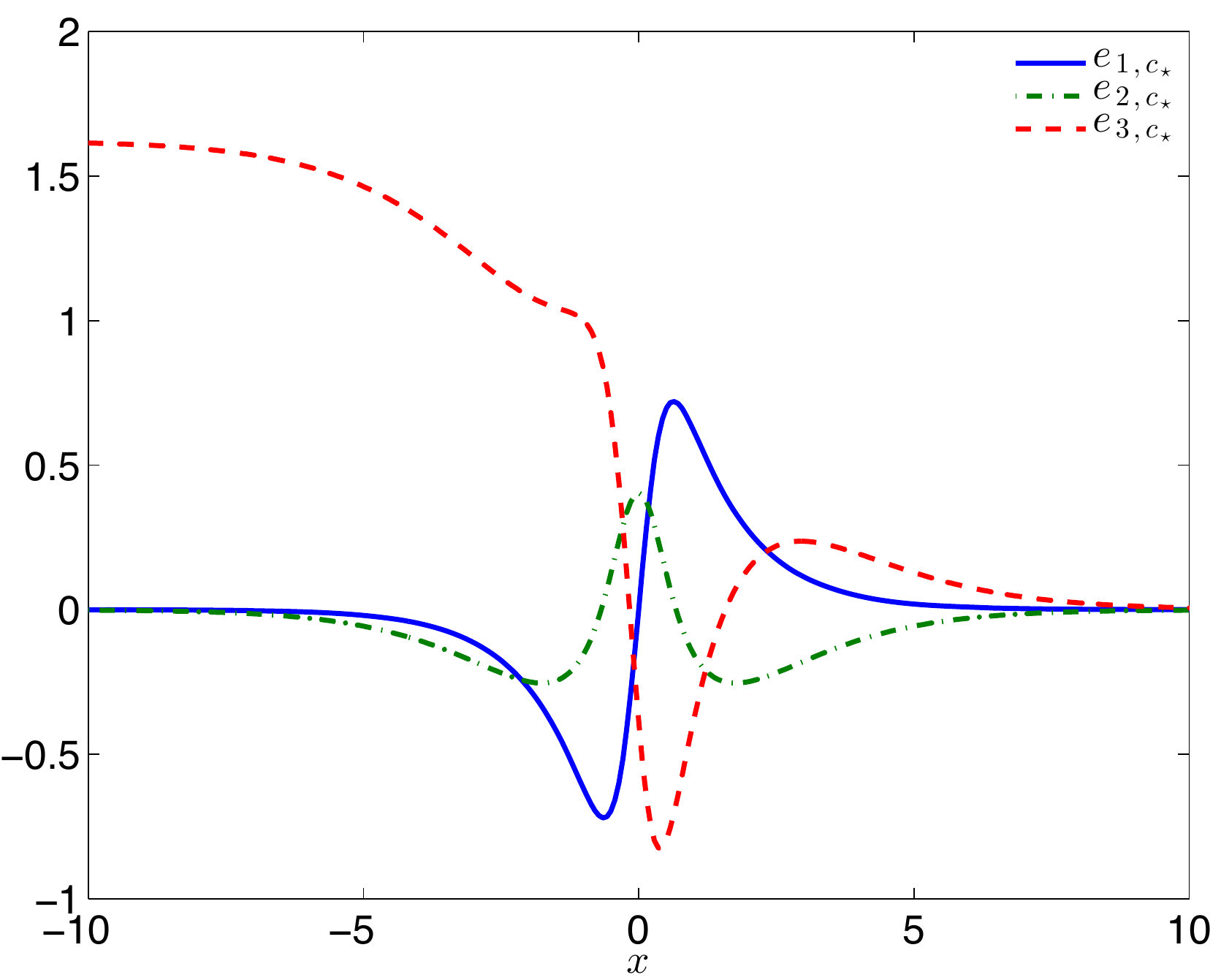}}
  \subfigure{\includegraphics[width=2.46in]{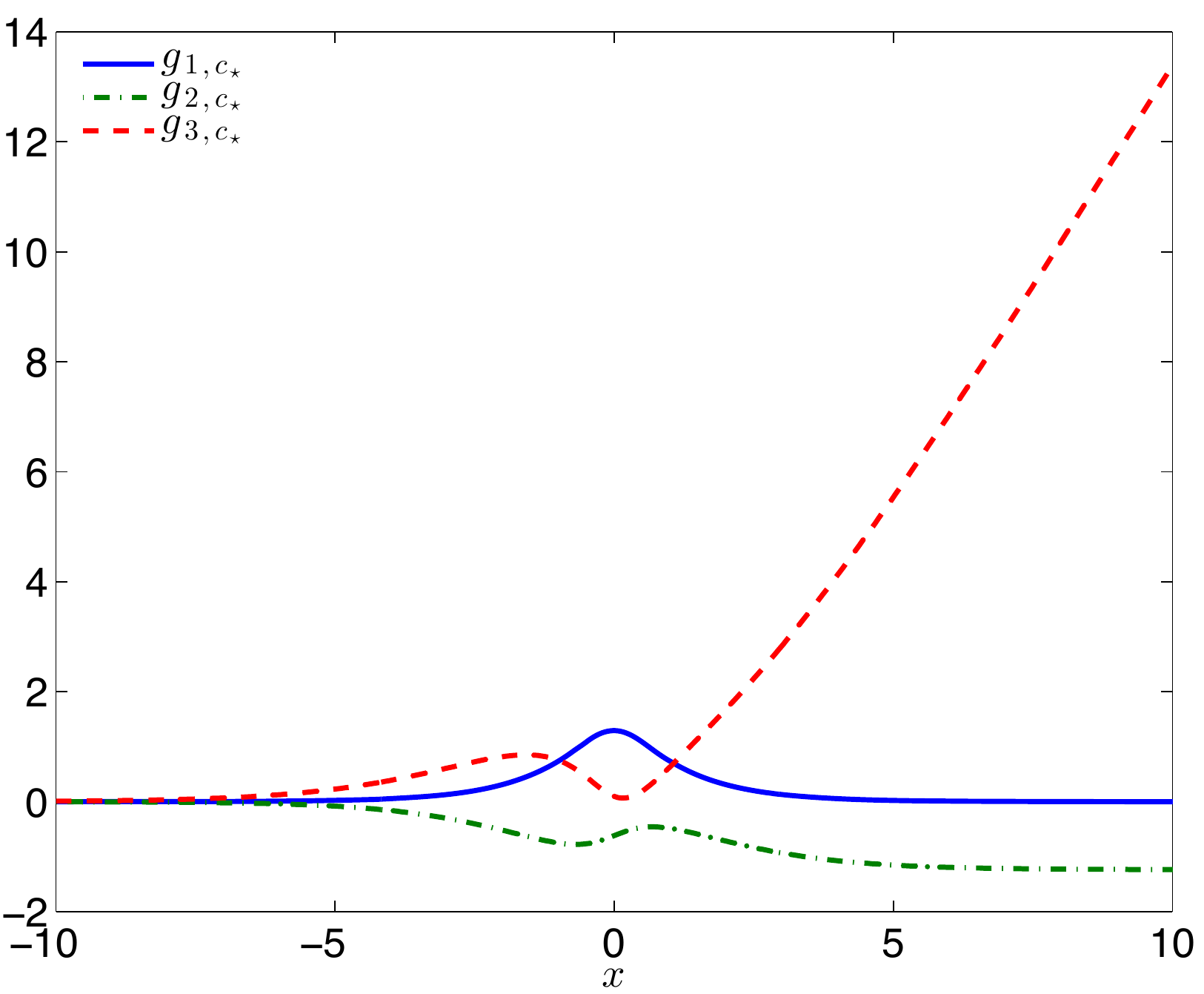}}
  \caption{The elements of the generalized kernels of $A_\cstar$ and
    $A_\cstar*$, computed with $M=500$ grid points.}
  \label{f:kernel}
\end{figure}

The kernel functions are given in Figure \ref{f:kernel}.  As discused,
$e_{3,\cstar}$, $g_{2,\cstar}$ and $g_{3,\cstar}$ are clearly not in
$L^2$.

\newpage
\bibliographystyle{abbrv}

\bibliography{sat_kdv}

\end{document}